\documentclass{svproc}

\usepackage{url}

\usepackage{cite}
\usepackage{dirtytalk}
\usepackage{amsmath,amssymb,amsfonts}
\usepackage{algorithmic}
\usepackage{graphicx}
\usepackage{float}
\usepackage{textcomp}
\usepackage{xcolor}
\usepackage{listings}
\usepackage{comment}
\usepackage{array}
\usepackage{stfloats}

\begin{document}
\mainmatter              % start of a contribution
\title{Resilient Risk based Adaptive Authentication and Authorization (RAD-AA) Framework \thanks{Preprint edition \textit{Nov 2022}}}
\titlerunning{RAD-AA Framework}  % abbreviated title (for running head)
\author{Jaimandeep Singh\inst{1\thanks{Email: jaimandeep.phdcs21@nfsu.ac.in   \hfill ORCID: 0000-0001-6266-1275}} \and Chintan Patel\inst{2} \and
Naveen Kumar Chaudhary\inst{1}}
\authorrunning{J Singh et al.} % abbreviated author list (for running head)
% \tocauthor{Ivar Ekeland, Roger Temam, Jeffrey Dean, David Grove,
% Craig Chambers, Kim B. Bruce, and Elisa Bertino}
%
\institute{National Forensic Sciences University, Gandhinagar, Gujarat, India \\
\and
The University of Sheffield, Sheffield, UK}

\maketitle              % typeset the title of the contribution

\begin{abstract}
In recent cyber attacks credential theft has emerged as one of the primary vectors of gaining entry into the system. Once attacker(s) have a foothold in the system, they use various techniques including token manipulation to elevate the privileges and access protected resources. This makes authentication and token based authorization a critical component for a secure and resilient cyber system. In this paper we discuss the design considerations for such a secure and resilient authentication and authorization framework capable of self-adapting based on the risk scores and trust profiles. We compare this design with the existing standards such as \textit{OAuth 2.0}, \textit{OIDC} and \textit{SAML 2.0}. We then study popular threat models such as \textit{STRIDE} and \textit{PASTA} and summarize the resilience of the proposed architecture against common and relevant threat vectors. We call this framework as \textit{Resilient Risk based Adaptive Authentication and Authorization (RAD-AA)}. The proposed framework excessively increases the cost for an adversary to launch and sustain any cyber attack and provides much-needed strength to critical infrastructure. We also discuss the machine learning (ML) approach for the adaptive engine to accurately classify transactions and arrive at risk scores.
% We would like to encourage you to list your keywords within
% the abstract section using the \keywords{...} command.
\keywords{Federated Authentication, Delegated Authorization, Cyber Resilience, Adaptive Engine, Identity Management Systems, Threat Models, secure architecture and framework, OAuth 2.0, OpenID Connect, SAML 2.0}
\end{abstract}
\section{Introduction}
\noindent As per the July-2022 report by IBM \cite{IBM2022}, one of the most frequent reasons for a data breach is using stolen or compromised credentials. The primary attack vector in 19\% of incidents was stolen or exposed credentials. Stolen credentials can also lead to significant damage to the complete ecosystem as was seen in the case of ransomware attack on the colonial pipeline system in the United States \cite{su2021ransomware}. Nowadays hackers do not break into the system but log into it, making the authentication and authorization systems a core design feature of any security architecture. With the help of authentication, we decide \say{who can enter} and with the help of authorization, we decide \say{up to what level he or she can access the system}.

Risk and trust based adaptive approach for authentication and authorization (RAD-AA) is required for a secure and resilient cyber system. Stringent security requirements are known to increase the friction in user experience. The adaptive approach also helps in reducing this friction by adjusting security requirements based on the context and the flow. In a traditional authentication system the credentials of a user are verified in a sequential manner by employing various means such as user name, password, one-time password or biometrics. However, in RAD-AA approach the credential verification requirements are dependent on the \textit{risk score} of the transactions and the \textit{trust relationship} between different entities of the ecosystem. Example: For a high risk transaction, the system may ask for MFA such as approving the notification received over the device or answer the security questions, while at low risk, the system may continue based on just user name and password. RAD-AA based framework decides \textit{risk score} based on the user's or system's state. For example: If a user tries to log in from a different geographic location, device, vulnerable application, or browser, then the system increases the \textit{risk score} for the transaction. Sometimes, if there is an attack on the complete critical infrastructure of the organization, then the RAD-AA framework-based system increases the \textit{risk score} for the complete system and expects that each user must pass through the high-security verification. 
Similarly, for access control or authorization, based on \textit{risk score}, the RAD-AA system can upgrade, degrade or revoke the rights given to a user or group of users. In this paper, we propose a novel Resilient Risk based Adaptive Authentication and Authorization (RAD-AA) framework that is attack-resilient and highly adaptive to the underlying risk factors. 

%The RAD-AA framework based system can define a dynamic access policy so that for any authenticated user, it can perform the activity monitoring and maintain the \textit{personal trust score}, which can then be used to calculate \textit{personal trust score}, \textit{user risk score} and current \textit{critical infrastructure risk score}. The RAD-AA system can then make dynamic decisions on the cumulative risk score.   

% \begin{figure}[H]
%     \centering
%     \includegraphics[scale=0.27]{Elements.pdf}
%     \caption{Elements of Cyber Resilience \cite{}}
%     \label{fig:1}
% \end{figure}

The remainder of the paper is as follows: Section \ref{sec2:relatedwork} presents related work. In this section, we discuss the existing features of \textit{OAuth 2.0}, \textit{OpenID connect}, and \textit{SAML 2.0} standards. The most common threat models are discussed in section \ref{sec3:ThreatModel} followed by design considerations for RAD-AA framework in section \ref{sec4:ProposedFramework}. The detailed architecture and threat matrix of RAD-AA framework is discussed in section \ref{sec5:proposedarchitecture}. In Section \ref{sec:ml-class} we discuss ML based approach for adaptive engine. This is followed by conclusion and future work in section \ref{sec6:conclusionandfuturework}.  

% Critical infrastructure cyber resilience
% [To be completed by Chintan]
%\subsection{Adaptive Systems}
% if possible spray some maths on adaptive systems
\section{Related Work}
\label{sec2:relatedwork}
There are several authentication and authorization frameworks available and adopted by the industries. \textit{OAuth 2.0} \cite{OAuth-2.0-rfc6749} protocol provides authorization for mobile, web and desktop applications and other smart devices. The OAuth 2.0 framework consists of four essential components: \textit{resource owner, resource server, client, authorization framework}. The RFC 6819 \cite{OAuth-2.0-Threat-rfc6819} defines the attacker capabilities and threat model for the \textit{OAuth 2.0} standard such as obtaining client secrets, obtaining refresh tokens, obtaining access tokens, phishing of end-user credential, open redirectors on client, password phishing.

Since, \textit{OAuth 2.0} provides only authorization framework, \textit{OpenID connect} \cite{sakimura2014openid} integrates an identity layer to it and enables authentication of the end user for the client side applications. An \textit{OpenID connect} permits all types of clients such as JSClient, web client, and mobile client.      

An open XML-based \textit{SAML 2.0} (standard called Security Assertion Markup Language) \cite{gross2003security,  cantor2005metadata} is frequently used to exchange authentication and authorization (AA) data amongst federated organisations. With the help of \textit{SAML 2.0}, the end user can log into multiple web-applications using the same credentials. \textit{SAML 2.0} enables single sign-on (SSO) facilities to access several independent applications with the help of an identity provider. The \textit{SAML 2.0} provides numerous advantages, such as users need not remember multiple user names and passwords, which also reduces access time, cost reduction, and labour cost reduction. 

There are other several authentication protocols such as \textit{LDAP} (Lightweight Directory Access Protocol), \textit{Kerberos}, \textit{RADIUS} but considering industry adoption and need of resilient adaptive authentication and authorization framework, we have compared the proposed framework with \textit{OAuth 2.0}, \textit{OpenID connect}, and \textit{SAML 2.0} in section \ref{sec4:ProposedFramework}.

\section{Threat Models}
\label{sec3:ThreatModel}
This section presents widely used threat models which are relevant to our proposed framework.  Threat modelling is the security procedure used to identify, classify, and examine potential risks. Threat modelling can be carried out either proactively during design and development or resolutely after a product has been released. In either situation, the method identifies the possible harm, the likelihood that it will happen, the importance of the issue, and the ways to remove or lessen the threat. Threats and vulnerabilities are frequently paired in order to find the risk to an organization or a system. 

In this paper, we have adopted the relevant portions of the existing threat models and tailored them based on the characteristics of different entities of the framework and their interactions with internal and external entities. The first threat model we have considered is STRIDE, developed by Microsoft \cite{shostack2008experiences}. The STRIDE model discusses six different types of threats. The first threat is \textit{Spoofing} where the adversary spoofs user identity and tries to access the resources with \textit{secure authentication} as a desirable property. The next threat is \textit{Tampering} where an adversary tampers the messages communicated over the open channel, and the desired property is \textit{integrity}. The third threat is \textit{repudiation} where attacker performs some illegal activities and denies performing those activities. The desirable property to tackle this threat is \textit{non-repudiation}, achievable using secure signature methods. The next threat is \textit{information disclosure} where the adversary tries to read or access the secret information. The desirable property is the \textit{confidentiality} that can be achieved using secure encryption/decryption system. The next threat is \textit{denial of service attack} where an adversary tries to prevent an end-user from accessing services where the desirable property is \textit{availability} that is achieved through intrusion detection and intrusion prevention system. According to the STRIDE threat model, the last threat is \textit{elevation of privilege} where an attacker is either an insider or somehow became the trusted insider with the high privilege to destroy the system. In this paper we consider a number of threats from the STRIDE model such as \textit{spoofing identity} and \textit{tempering}.     

The \textit{PASTA} (Process for Attack Simulation and Threat Analysis)  is a risk-centric threat modelling technique that incorporates risk analysis and context into the complete security of critical infrastructure from the start \cite{ucedavelez2015risk}. The \textit{PASTA} allows threat modelling linearly through the interdependent seven stages. In the \textit{first stage}, \textit{PASTA} defines the \textit{objectives for risk analysis} and provides well-defined business objectives and analysis reports. In \textit{stage two}, \textit{PASTA} defines the \textit{technical scope} and tries to understand possible attack surface components and provides a detailed technical report on all attack surface components. In the \textit{third stage}, \textit{PASTA} performs \textit{application decomposition and analysis} and provides a data-flow diagram, interfaces list with trust level, asset list, and access control matrix. In the \textit{fourth stage}, \textit{PASTA} performs \textit{threat analysis} and generates a list of threat agents, attack vectors, incident event reports. In \textit{stage five}, it performs \textit{vulnerability assessments} and provides scoring based on CVSS (Common Vulnerability Scoring System). In stage six, \textit{PASTA} performs \textit{attack modelling} and simulation of various well-known threats. As an outcome of stage six, it provides attack trees and possible attack paths. In the last (\textit{seventh}) stage, it performs \textit{risk analysis and management} where it outcomes risk profile, risk mitigation strategy, and threat matrix. We have considered the \textit{PASTA} framework where ever required during framework design.    

\section{Design Considerations for RAD-AA Framework}
\label{sec4:ProposedFramework}
In this paper we propose a Risk based Adaptive Authentication and Authorization (RAD-AA) framework which is secure by design and excessively increases the cost for the attacker. The design considerations include adaptive ecosystem which is capable of modifying the security requirements and the access rights to the protected resources based on the risk score of the transactions and interactions between different entities of the ecosystem. Table \ref{table:1} presents comparison of the proposed framework with the existing standards. 

The adaptive design enables the framework to anticipate the attack. In case of a breach, the adaptive design is able to \textit{withstand} and \textit{constraint} the level of damage by revoking or restricting the access rights granted or by de-authenticating already authenticated users or increasing the security requirements of the transactions. The design considerations of RAD-AA are given below:

\begin{itemize}
\item \textbf{Adaptive Engine for Cyber Resilience}. The authentication and authorisation transactions and trust level between different entities in the ecosystem should be able to adapt based on the risk score such as Geo-location, impossible travel, IP reputation, and device information. The system should be able to enhance, reduce or completely revoke the entry into the ecosystem or access to the protected resources based on the risk score.
\item {\textbf{Federated Authentication}. It is a system in which two parties trust each other to authenticate the users and authorize access to resources owned by them. In a Federated Authentication or Identity management system the identity of a user in one system is linked with multiple identity management systems. It extends beyond a single organization wherein multiple organizations can agree to share the identity information and join the federation. Once the users' login into their organizations, they can use this federated identity to access resources in any other organization within the federation.}
\item {\textbf{Delegated Authorization}. A delegated authorization system can delegate access rights to the services or processes in the form of assertions or claims. Examples include end-user authorization delegation to a web or native application.}
\item {\textbf{Decoupling Authentication and Authorization}. Decoupled authentication and authorization systems allow for modularity and provide necessary interfaces to replace one system with another. 
}
\item {\textbf{Confidentiality and Non-repudiation of claims}. The confidentiality will ensure that the contents of the claims are not revealed to unauthorized parties. Non-repudiation will assure the recipient that it is coming from the identified sender and that the data integrity is assured.}

\item {\textbf{Audience Binding}. The primary aim of binding the token to the audience or the client to which it is issued is to prevent unauthorized entities from using the leaked or stolen tokens. The token, when issued, is bound to a public key of the audience or the client to which it is issued. The client now needs the corresponding private key to use the token further. This will protect against the token's misuse by unauthorized parties that do not possess the private key. It provides an added assurance to the receiver that such a claim token has been sent by the sender authorized to use it.}
\item {\textbf{Trust relationships}. The system should be capable of establishing trust relationships between different entities of the ecosystem. The trust relationship will also depend on how the two entities identify each other and what is the relationship between the entities. The risk score of the client applications can be taken into consideration while deciding the trust relationship level.
}

\item {\textbf{Propagation of identity and claims between different ecosystems}. The requested resource or a subset of the resource may be required to be fetched from a different ecosystem. The necessitates the propagation of identity and claims across different ecosystems. When transitioning from one ecosystem to another, properties of confidentiality, non-repudiation, limiting the information contained in the claim and assurance of use by the authorized sender should be maintained.
}
\item {\textbf{Time-limited Validity and Revoking issued claims}. The lifetime of the issued tokens should be limited to avoid misuse of the stolen tokens \cite{OAuth-2.0-rfc6749}. The ability to revoke the tokens should be in-built into the design.
}
\item {\textbf{Support for REST API architecture}. Most of the developers have now moved from WS-* to REST as conceptualized by Fielding in his seminal PhD dissertation \cite{fielding2000architectural} and stateless APIs \cite{masse2011rest}. The system design should therefore support the REST API architecture. The REST can simply be described as HTTP commands pushing JSON packets over the network as defined in RFC 8259 \cite{rfc8259}.
}
\end{itemize}

\begin{table}[H]
\begin{center}
\caption{Comparison of existing standards with proposed design considerations of RAD-AA}
\begin{tabular}{ | m{12em} | m{6em}| m{8em} | m{6em} | m{5em} |} \hline
 \textbf{\textit{Design Considerations}} & \textbf{\textit{OAuth 2.0}} \cite{OAuth-2.0-rfc6749} & \textbf{\textit{OpenID Connect}} \cite{sakimura2014openid} & \textbf{\textit{SAML 2.0}} \cite{gross2003security} & \textbf{\textit{Proposed Framework}} \\ 
 \hline
 \textbf{Authentication} & NO & YES & YES & YES \\  \hline
 \textbf{Adaptive Engine for Cyber Resilience} & NO & NO & NO & YES \\  \hline
 \textbf{Federated Authentication} & NO & SSO Only & SSO Only & YES \\ \hline
 \textbf{Delegated Authorization} & YES & YES & YES & YES \\  \hline
 \textbf{Decoupling Authentication and Authorization} & YES & YES & YES & YES \\  \hline
 \textbf{Out of the box support for Confidentiality and Non-repudiation of claims} & NO & NO & NO & YES \\ \hline
 \textbf{Audience Binding} & NO & NO & NO & YES \\ \hline
 \textbf{Trust Relationships} & NO & NO & NO & YES \\  \hline
 \textbf{Time limited validity of claims} & YES & YES & YES & YES \\  \hline
 \textbf{Ability to revoke issued tokens} & NO & NO & NO & YES \\  \hline
 \textbf{Support for REST API architecture} & YES & YES & YES & YES \\
 \hline
 \textbf{Extensible to ML Based Classification} & NO & NO & NO & YES \\
 \hline
\end{tabular}
\label{table:1}
\end{center}
\end{table}
 
%which basically includes trust boundaries, Dataflow Paths, Input Points, Privileged Operations, Details about Security Stance and Approach

\section{Architecture and Threat Matrix for RAD-AA Framework}
\label{sec5:proposedarchitecture}
\begin{figure}[!ht]
    \centering
    \includegraphics[scale=0.3]{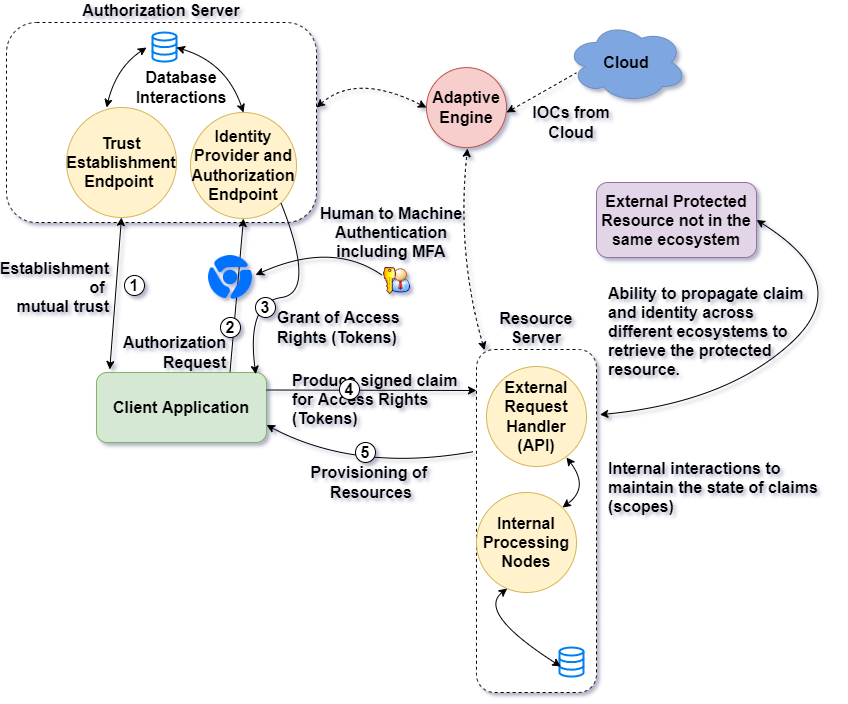}
    \caption{Architecture of RAD-AA framework}
    \label{fig:2}
\end{figure}

In this section we describe the detailed architecture of RAD-AA based on the various existing standards \cite{OAuth-2.0-rfc6749}\cite{gross2003security} \cite{sakimura2014openid}\cite{rfc8259}\cite{rfc9126}, threat considerations \cite{OAuth-2.0-Threat-rfc6819}, best current practices \cite{ietf-oauth-security-topics-20}\cite{rfc8705} and augmented features of OAuth 2.0 \cite{singh2022oauth}. The architecture, entities, protocol flow and their interactions is given at Figure \ref{fig:2}.

We have analyzed various threats that can manifest in any authentication and authorization protocol. We have then brought out the features in the proposed framework that can mitigate the effect of these threats by anticipating and adapting itself in face of changing operating environment.
%Fonts in the figure are to small. 

\subsection {{Entities}}
The RAD-AA framework consists of following entities:
\begin{itemize}
\item \textbf{\textit{Resource Owner}}. It is an entity which is capable of granting access to a protected resource. \end{itemize}
\begin{itemize}
\item \textbf{\textit{Resource Server (RS)}}. This server hosts the protected resources and is capable of accepting and responding to protected resource requests based on the access tokens and scopes. 
\end{itemize}
\begin{itemize}
\item \textbf{\textit{Client}}. This is an application which makes request for accessing protected resource requests on behalf of the resource owner after due authorization by the resource owner. 
\item \textbf{\textit{Authorization Server (AS)}}. This server issues access tokens to the client after successfully authenticating the resource owner and obtaining authorization.
\item \textbf{\textit{Adaptive Engine}}. This engine analyses various transactions in the protocol flow. It then assigns the risk score which is used by various entities like authorization server, resource server to modify their behaviour by making the security requirements more stringent or to limit or deny access to the protected resources.
\end{itemize}

\subsection {{Protocol Flow and Threat Matrix}}

The summary of common threat vectors and resilience features of RAD-AA framework is given at Table \ref{table:2}.

\begin{table}[!ht]
\begin{center}
\caption{Summary: Threat Matrix for RAD-AA Framework}
\begin{tabular}{| m{12em} | m{25em}|} \hline
 \textbf{\textit{Threat Vectors}} & \textbf{\textit{RAD-AA Cyber Resilience Features}} \\  \hline
 \textbf{Client Impersonation} & Mutual authentication using mTLS \cite{rfc8705} or DPoP \cite{ietf-oauth-dpop-10}. \\  \hline
 \textbf{Cross-Site Request Forgery (CSRF)} & AS supports Proof Key for Code Exchange (PKCE) \cite{rfc7636}. \\  \hline
 \textbf{Authorization Server (AS) Mix-Up attack} & Validation of the issuer of the authorization response. \\  \hline
 \textbf{Cross-Origin Resource Sharing (CORS)} & 
 \begin{itemize}
 \item Establish trust relationships through mutual authentication.
 \item Calculate risk score before allowing transactions.
 \item Information in HTTP request is assumed fake.
 \end{itemize}\\  \hline
 \textbf{Cross Site Scripting (XSS)} & 
 \begin{itemize}
 \item CSP headers.
 \item Input validation.
 \end{itemize} \\ \hline
 \textbf{DDoS on AS} &
 \begin{itemize}
 \item Check HTTP request parameters are not pointing to unexpected locations, RFC 9101 \cite{rfc9101}.
 \item Adaptive engine to thwart malicious requests based on the risk score and the trust relations between the client applications and the authorization server.
 \end{itemize} \\  \hline
 \textbf{Access Token Injection} & Audience restricted token binding. \\ \hline
 \textbf{Access Token Replay} & Sender-constrained and audience-restricted access tokens. \\ \hline
\end{tabular}
\label{table:2}
\end{center}
\end{table}

\subsubsection {\textbf{Risk Score Based Adaptive Engine\newline}}
The adaptive engine will determine the risk score of each transaction in the protocol flow based on AI/ML models. It will provide its inputs to all the entities in the ecosystem such as authorization server and the protected resource server. The engine will take into consideration various parameters such as the level of trust relation between the sender and the receiver,  Geo-location, impossible travel, IP reputation, and device information and then classify the transaction into \textit{LOW}, \textit{MEDIUM} and \textit{HIGH} risks. 

The authorization server and the protected resource server based on the classification of the transaction can do either of the following:

\begin{itemize}
\item \textbf{\textit{Raise the security requirements}}. The security requirements of the transaction can be made more stringent by demanding the sender to provide additional verification details.
\end{itemize}
\begin{itemize}
\item \textbf{\textit{Accept or limit/Reject the Request}}. The AS/RS based on the risk score of the transaction can either accept the transaction or lower the authorization of the scopes or completely reject the transaction as a high risk transaction.
\end{itemize}

The authorization and access to the protected resources will still be governed by the requirements specified in the protocol. The engine will only classify the transaction cost as an additional security layer. The engine classification at no time will bypass the requirements of other validation requirements as given in the specifications. The common risks associated with the adaptive engine which are ordinarily associated with AI/ML engines must be taken into consideration\cite{mclean2021risks}.

\subsubsection {\textbf{Establishment of Mutual Trust\newline}}
The client applications like native desktop or mobile app, JavaScript based Single Page App or the web server based web apps need to establish a level of trust while authenticating their identities with the AS/RS. Each such client authentication will be assigned a trust assurance level which will regulate the ability of the client application to acquire elevated authorization rights or scopes for an extended duration.
\begin{itemize}
\item {\textbf{Trust Assurance Level 0}. This is the minimum trust assurance level. The application does not have means of mutual identification and authentication. The delegated authorization rights would be restricted to a minimum level of permissions (scopes). The lifespan of the delegated authorization rights (in terms of access and refresh tokens) will be minimal or for one time use only.  
}

\item {\textbf{Trust Assurance Level 1}. client application has the means of establishing mutual identification and authentication by using mTLS\cite{rfc8705} or DPop\cite{ietf-oauth-dpop-10}. The delegated authorization rights/scopes will be more permissive in nature with an extended lifetime validity.
}
\end{itemize}

\smallskip
\textit {{Threat Vector:}} Client Impersonation.

\textit {{Threat Description and Artifacts:}} The identity of the client application can be spoofed and a malicious client can impersonate as a genuine client.

\textit {{Mitigating Features:}} In the proposed framework the client and authorization server needs to establish a trust relation by mutually identifying and authenticating each other by using mTLS \cite{rfc8705} or DPoP \cite{ietf-oauth-dpop-10} or similar such standards.

\subsubsection {\textbf{Authorization Request}\newline}
The client application will initiate the request to the authorization server for grant of access rights to acquire the protected resources owned by the resource owner. The resource owner will authorize the scopes that the authorization server should grant to the requesting client. The access rights of the client will be constrained and restricted to the scopes authorized by the resource owner. The authorization server will first verify the identity of the resource owner by passing the request to the federated identity management system \cite{Hedberg2022openid}. The Pushed Authorization Requests (PAR) as defined in RFC 9126 \cite{rfc9126} will be used to initiate the authorization request by the client.

\smallskip
\textit {Threat Vector:} Cross-Site Request Forgery (CSRF).

\textit {Threat Description and Artifacts:} Clients Redirect / Request URIs are susceptible to CSRF.

\textit {Mitigating Features:} The framework prevents CSRF attacks against client's redirect / request URIs by ensuring that the authorization server supports Proof Key for Code Exchange (PKCE)\cite{rfc7636}. The CSRF protection can also be achieved by using \texttt{"nonce"} parameter, or \texttt{"state"} parameter to carry one-time-use CSRF tokens.

\smallskip
\textit {{Threat Vector:}} Authorization Server (AS) Mix-Up attack.

\textit {{Threat Description and Artifacts:}} The AS attack can manifest when multiple AS are used in which one or more is a malicious AS operated by the attacker.

\textit {{Mitigating Features:}} The framework prevents mix-up attacks by validating the issuer of the authorization response. This can be  achieved by having the identity of the issuer embedded in the response claim itself, like using the \texttt{"iss"} response parameter.

\smallskip
\textit {{Threat Vector:}} Cross-Origin Resource Sharing (CORS) attacks.

\textit {{Threat Description and Artifacts:}} CORS allows web applications to expose its resources to all or restricted domains. The risk arises if the authorization server allows for additional endpoints to be accessed by web clients such as metadata URLs, introspection, revocation, discovery or user info endpoints. These endpoints can then be accessed by web clients.

\textit {{Mitigating Features:}} In the proposed framework we establish trust and calculate risk score before allowing any transactions. It is accepted that access to any of the resources as all the information contained in the HTTP request can be faked.

\smallskip

\textit {{Threat Vector:}} Cross Site Scripting (XSS) attacks.

\textit {{Threat Description and Artifacts:}} Cross-Site Scripting (XSS) attack risk arises when the attacker is able to inject malicious scripts into otherwise benign and trusted websites.

\textit {{Mitigating Features:}} Best practices like injecting the Content-Security-Policy (CSP) headers from the server are recommended, which is capable of protecting the user from dynamic calls that will load content into the page being currently visited. Other measures such as input validation are also recommended.

\smallskip

\textit {{Threat Vector:}} DDoS Attack on the Authorization Server.

\textit {{Threat Description and Artifacts:}}  A large number of malicious clients can simultaneously launch a DoS attack on the authorization server by pointing to the \texttt{"request\_uri"} as defined in PAR RFC 9126 \cite{rfc9126}.

\textit {{Mitigating Features:}} To mitigate the occurrence of such an attack, the server is required to check that the value of the \texttt{"request\_uri"} parameter is not pointing to an unexpected location as recommended in the RFC 9101 \cite{rfc9101}. Additionally, the framework employs an adaptive engine to thwart such requests based on the risk score and the trust relations between the client applications and the authorization server.

\subsubsection {\textbf{Delegation of Access Rights \newline}}
Once the resource owner has authorized the request, authorization grant has been received by the resource owner and the request received from the client application is valid, the authorization server issues access tokens and optional refresh tokens as defined in RFC 6749 \cite{OAuth-2.0-rfc6749}.

\smallskip
\textit {{Threat Vector:}} Access Token Injection.

\textit {{Threat Description and Artifacts:}} In this attack, the attacker attempts to utilize a leaked access token to impersonate a user by injecting this leaked access token into a legitimate client as defined in draft OAuth 2.0 Security Best Current Practice, 2022 \cite{ietf-oauth-security-topics-20}.

\textit {{Mitigating Features:}} In the proposed framework, the token is issued after binding to the client application.

\subsubsection {\textbf{Accessing Protected Resources\newline}}
The client application can access protected resources by presenting the access token to the resource server as defined in OAuth 2.0 RFC 6749 \cite{OAuth-2.0-rfc6749}. The resource server will then check the validity of the access token. It will also ensure that the token has not expired and that the requested resource is conformance with the scopes authorized in the token.

\smallskip
\textit {Threat Vector:} Access Token Replay Attack.

\textit {Threat Description and Artifacts:} An attacker can attempt to replay a valid request to obtain or modify/destroy the protected resources as defined in OAuth 2.0 threat model RFC 6819 \cite{OAuth-2.0-Threat-rfc6819}.

\textit {{Mitigating Features:}} The proposed framework mandates sender-constrained and audience-restricted access tokens as defined in draft OAuth 2.0 Security Best Current Practice, 2022 \cite{ietf-oauth-security-topics-20}. In addition, the resource server may reduce the scope or completely deny the resource based on the risk score of the client application from where the request has materialized.

\section{ML based Classification approach for Adaptive Engine}
\label{sec:ml-class}
The adaptive engine can employ various ML algorithms or models to classify the transaction data into \textit{HIGH}, \textit{MEDIUM} or \textit{LOW} risk on a real-time basis. Though there is a high chance that data received by the adaptive engine can be noisy, the K-Nearest Neighbour (KNN) \cite{peterson2009k} or Random Forest (RF) \cite{biau2016random} can be adopted based on their capability to work with multidimensional feature sets. RF requires a large training dataset that can be gradually built into the cloud based system and later used. Adoption of \textit{Incremental Machine Learning (IML)} \cite{solomonoff2002progress} approaches over the adaptive engine can provide more accurate classification as it continuously trains the model and over a period of time.

The output of the adaptive engine is taken into consideration by various systems to grant or deny permissions or accept/reject or terminate the transaction, and restart the process. Some important features that can provide a high degree of correlation to classify the transaction data are tabulated in Table \ref{tab:CF}.
\begin{table}
 \caption{Classification Features}
    \centering
    \begin{tabular}{|p{8em}|p{21em}|} \hline
    \textbf{Features} & \textbf{Descriptions} \\ \hline
       IP reputation  & The IP reputation will let the system know if the request is coming from an IP with a bad reputation. \\ \hline
     Geo Location
   & This will identify the geo location of the transaction.
\\ \hline
      Impossible Travel
  & This feature will indicate if the same user has connected from two different countries or geographical locations and the time between these locations can't be covered through conventional travel means.
\\ \hline
    Device Info
    & Device information will provide the necessary data about the device being used by the client application.
 \\ \hline
      NIDS output
  & It will take Netflow data as input stream and classify data as malicious or benign.
\\ \hline
   Trust Assurance Level
     & Trust Assurance Level will determine the degree of trust that server can have on the client application based on the verifiable client identity.
\\ \hline
    \end{tabular}
    \label{tab:CF}
\end{table}
\begin{figure}
    \centering
    \includegraphics[width=\linewidth]{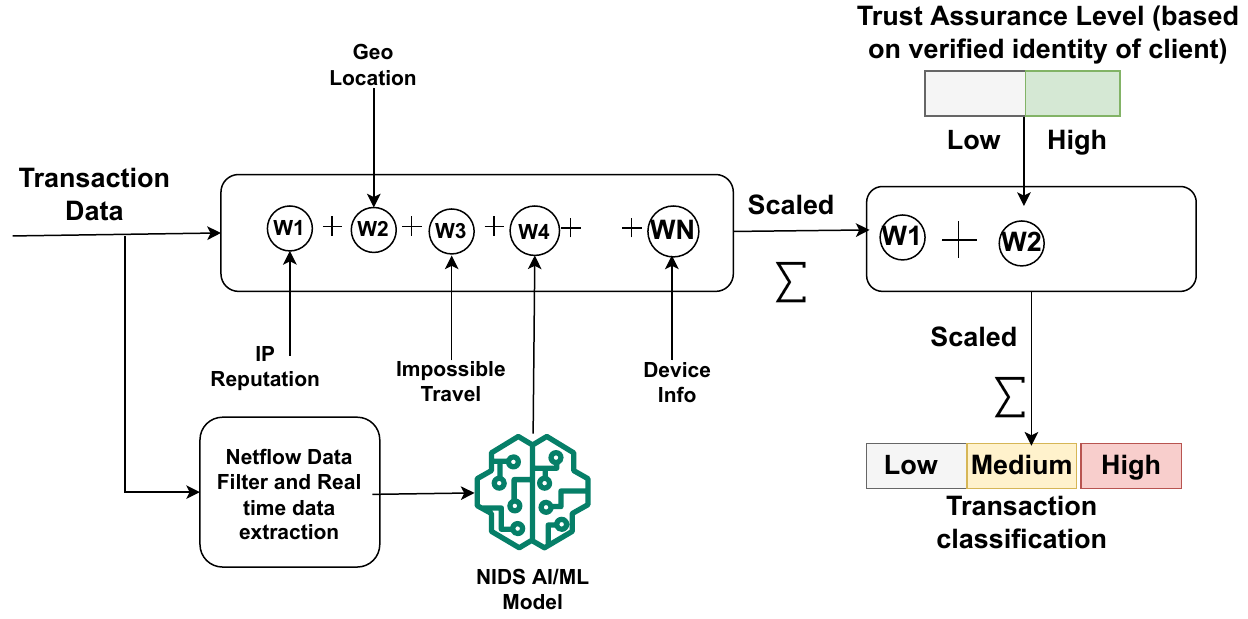}
    \caption{Use of ML Based Classification in Adaptive Engine}
    \label{fig:ML_Classificaion}
\end{figure}

The Network Intrusion Detection System (NIDS) will take Netflow data as an input stream. The output of NIDS will become one of the input features for the adaptive engine. Trust Assurance Level is an important feature based on the client applications' verifiable identity and authentication. A client application that can authenticate itself by using various technologies/protocols like mutual-TLS will have a higher trust assurance level than client applications that cannot authenticate themselves. The model or the algorithm can be fine-tuned by adding other features, hyper-parameters or removal/addition of necessary weights or biases for some features over others.

\section{Conclusion and Future work}
\label{sec6:conclusionandfuturework}
This paper presents the design considerations and architecture for a Resilient Risk-based Adaptive Authentication and Authorization (RAD-AA) Framework. The proposed framework achieves higher level of resilience than the existing frameworks such as \textit{OAuth 2.0}, \textit{OpenID Connect} and \textit{SAML 2.0} with the help of adaptive engine based on risk score and trust relationship. In future, we aim to continue developing this framework and consider further aspects related to its deployment on different platforms using ML.

\bibliographystyle{splncs03}
\bibliography{references}
\end{document}